# Experimental assessment of clinical MRI-induced global SAR distributions in head phantoms

J. Blackwell[1,2*], G. Oluniran[1*], B. Tuohy[3], M. Destrade[2], M. J. Kraśny[1†], N. Colgan[1†]

1 School of Physics, National University of Ireland Galway, Galway, Ireland

2 School of Mathematics, Statistics and Applied Mathematics, National University of Ireland Galway, Galway, Ireland

3 Medical Physics and Bioengineering, Galway University Hospital, Galway, Ireland

## ABSTRACT

**Objective:**
Accurate estimation of SAR is critical to safeguarding vulnerable patients who require an MRI procedure. The increased static field strength and RF duty cycle capabilities in modern MRI scanners mean that systems can easily exceed safe SAR levels for patients. Advisory protocols routinely used to establish quality assurance protocols are not required to advise on the testing of MRI SAR levels and is not routinely measured in annual medical physics quality assurance checks. This study aims to develop a head phantom and protocol that can independently verify global SAR for MRI clinical scanners.

**Methods:**
A four-channel birdcage head coil was used for RF transmission and signal reception. Proton resonance shift thermometry was used to estimate SAR. The SAR estimates were verified by comparing results against two other independent measures, then applied to a further four scanners at field strengths of 1.5 T and 3 T.

**Results:**
Scanner output SAR values ranged from 0.42-1.52 W/kg. Percentage SAR differences between independently estimated values and those calculated by the scanners differed by 0-2.3%.

**Conclusion:**
We have developed a quality assurance protocol to independently verify the SAR output of MRI scanners.

# 1. INTRODUCTION

During a magnetic resonance imaging (MRI) procedure, most of the transmitted radiofrequency (RF) energy is transformed into heat by the induction of eddy currents in the body and deposited into the patients' tissue. Specific energy absorption rate (SAR) expressed as watts per kg [W/kg] is the mass normalised dose metric for measuring the rate of energy deposition to tissue [1].

Pennes' bioheat equation defines temperature distribution within biological tissues and incorporates effects from metabolism, perfusion and thermal conduction [2]. Depending on the pattern of RF radiation absorption and the efficiency of heat regulatory mechanisms in body tissues, RF heating can cause adverse biological effects [1,2,3]. Thus, accurate estimation of SAR is critical to safeguarding unconscious/sedated patients, patients with compromised thermoregulation, implant patients, pregnant patients and neonates who require an MRI procedure.

The EU standardization has mandated that all scanners must measure SAR in patients and develop system safeguards to ensure that the limits (IEC60602-3-33) are not exceeded. The standard specifies derived SAR limits of 2 W/kg averaged over the whole-body (wbSAR), 3.2 W/kg for the head (hdSAR), and 2–10 W/kg for parts of the body (pbSAR) for normal controlled operating mode. At 1.5 T, most clinical MR imaging operates well below these regulatory thresholds but, because of the approximately quadratic SAR dependence on the strength of the main field, they pose a more frequent concern for 3 T acquisitions [4]. Ultra-high field MRI at 7 T and higher pose additional safety concerns [5].

Commercial MRI scanners provide an estimated SAR level for each scan calculated using factory-determined parameters. The accuracy of these values can vary over time and can be highlighted in cases where patients sustain RF burns [6]. To ensure the predictive methods are accurate, it is necessary to compare these calculations with experimental results. The need for SAR level validation was reinforced in August of 2015, when GE released an urgent medical device correction where 756 of their systems underestimated the head SAR delivered to patients [7]. When performing head or neck scans, the displayed SAR values could be lower than the actual SAR as predicted by SAR modelling and exceed the limit of 3.2 W/kg for some scans.

To estimate MRI SAR values, both numerical and experimental methods have been used. Extensive work has been carried out in the area of numerical SAR calculations. A detailed review of the various numerical methods and studies has been compiled by Hartwig [8]. Head SAR specific numerical calculations have also been investigated by Van Lier et al. and Wang et al. explored the physiological response of the head using the Pennes' bioheat equation [2,3]. Numerical simulations are inexpensive and can be done rapidly; however, it is also important to independently verify these results experimentally. There have been fewer experimental assessments of MRI SAR, with some giving global SAR values [9,10] and others attempting to obtain the more difficult unaveraged or 'local' SAR values [11-13]. No experimental assessments for head-averaged SAR values have been described in the literature to date and we believe the recent device correction highlights the need for developing a clinical phantom and experimental protocol.

Using short gradient echo (GRE) acquisitions of less than 11 seconds pre- and post-heating, we create a phase map from which we can determine the thermal shift due to temperature rise. We describe a method to experimentally determine the thermal shift in a $T_1$ doped MR phantom using proton resonance shift thermometry where the only source of heat is the radiofrequency fields produced by the imaging coils. From this thermal shift we can generate an estimate of the whole-body or "global" SAR weighted for a head mass. We then verify the results using independent measurements.

# 2. THEORY

## 2.1 Specific Absorption Rate (SAR)

The Pennes' bioheat equation (Eq. 1) is the standard model used to predict temperature distributions in living tissue [2].

$$\rho_t c_t \frac{\Delta T}{\Delta t} = \underbrace{\nabla.(k_t \nabla T)}_{Conduction} - \underbrace{W_b c_b (T - T_a)}_{Arterial\ blood\ heating} + SAR\rho_t \quad (1)$$

where $\Delta T$ is the change in temperature, $\Delta t$ is the period of heating, $\rho_t$ is the tissue density, $c_t$ is the specific heat capacity of the tissue, $k_t$ is the tissue thermal conductivity and SAR is the absorbed energy in watts per kilogram. SAR is equivalent to the heat created by the electric field within the tissue (Eq. 2).

$$SAR = \frac{\sigma}{\rho} E^2 [W/kg] \quad (2)$$

where $\sigma$ is the conductivity of the tissue [S/m], $\rho$ is the density of the tissue [kg/m] and E is the peak electric field (rms) [V/m]. A study by Wang et al. [3] concluded that for head-average SAR values of 3.0 W/kg or less, the effect of temperature-induced physiological changes is negligible (0.01–0.02 $°C$) [3]. Moreover, according to Shellock [1], there is an insignificant contribution from thermal conduction in SAR assessment. As our phantom is nonperfused and the period of heating is relatively short (~10 minutes), physiological changes can be ignored, reducing the SAR estimation based on heat capacity and temperature change over time (Eq. 3).

$$SAR \approx c_{phantom} \frac{\Delta T}{\Delta t} [W/kg] \quad (3)$$

where $c_{phantom}$ is the specific heat capacity of the phantom [J/kg.K], $\Delta T$ is the change in temperature and $\Delta t$ is the period of heating.

Temperature changes can be determined using a MR thermometry technique and there are several approaches including proton density, $T_1$ relaxation time of water protons, $T_2$ relaxation time of water protons, Diffusion - Brownian molecular motion and Proton resonance frequency shift of water protons. Below we shortly describe all of them with their usability in experimental SAR values estimation.

## 2.2 Proton density thermometry

Proton density depends linearly on the equilibrium magnetization $M_0$, which is determined by the Boltzmann distribution (Eq. 4).

$$PD \propto M_0 = \frac{N\gamma^2 I \hbar^2 (I+1) B_0}{3\mu_0 kT} = \chi_0 B_0 \quad (4)$$

where N is the number of spins per volume, $\gamma$ is the gyromagnetic ration, $\hbar$ is Plank's constant, $I$ is the quantum number of the spin system, $B_0$ is the magnetic flux density, $\mu_0$ is the permeability of free space, $k$ is the Boltzmann constant, $T$ is the absolute temperature of the sample and $\chi_0$ is the magnetic susceptibility.

As $M_0$ depends on the Boltzmann thermal equilibrium, relative temperature changes based on proton density-weighted images can be evaluated. This method requires long repetition times close to 10 seconds [14] which makes it unfeasible due to increased experimental error caused by the phantom cooling between successive acquisitions.

## 2.3 $T_1$ & $T_2$ relaxation time of water protons

The principle behind the temperature dependence of the $T_1$ relaxation time of water protons is that the $T_1$ relaxation time depends linearly on temperature [15,16] and can be described with equation 5:

$$T_1(T) = T_1(T_0) + m.(T - T_0) \quad (5)$$

where $T_0$ is the background temperature and $m = \frac{dT_1}{dT}$ is determined empirically for each type of tissue.

Qualitative temperature measurements can be acquired rapidly, but there are several difficulties. Firstly, the temperature dependence $m$ must be calculated for each type of tissue which makes imaging heterogeneous tissues more complex. This method also depends on the accuracy of measuring and extracting the $T_1$ relaxation time which will add to the experimental error.

Similarly, the $T_2$ relaxation time of water shows a temperature dependence. Also, the $T_2$ signal is decreased compared to pure water, reducing the accuracy of this method [17]. However, temperature measurements in adipose tissue have been demonstrated [18].

## 2.4 Diffusion - Brownian molecular motion

This method relies on the temperature dependence of the diffusion coefficient D which describes the thermal Brownian motion of molecules in a medium [19]. The temperature dependence of the diffusion coefficient is presented in equation 6:

$$D \approx e_a^{-\frac{E_a(D)}{kT}} \quad (6)$$

where $E_a(D)$ is the activation energy of the molecular diffusion of water, $k$ is the Boltzmann constant and $T$ is the absolute temperature.

If the temperature change $\Delta T$ is assumed to be small and that $E_a(D)$ is independent of temperature, the temperature change can be determined with equation 7.

$$\Delta T = T - T_0 = \frac{kT_0^2}{E_a(D)}\left(\frac{D-D_0}{D_0}\right) \quad (7)$$

This method has been used for in-vivo temperature recording. The benefit of this method is high sensitivity, but it has long acquisition times and high sensitivity to motion. The long acquisition time for this method could result in increased experimental error due to the phantom cooling between acquisitions.

## 2.5 Proton Resonance Frequency Shift (PRF) thermometry

In this work PRF was the chosen method to calculate the change in temperature of the phantom. The principle behind this method is described in detail by Rieke and Butts Pauly, Peters and Henkelman, and Olsrud et al. [14,20-22]. However, a general overview of this method is given below.

The local magnetic field $B_{local}$ determines the resonance frequency of a nucleus. The local magnetic field is related to the magnetic field $B_0$ of the scanner and is described by (Eq. 8)

$$B_{local} = (1 - \sigma_{total})B_0 + \delta B_0 \quad (8)$$

where $\sigma_{total}$ is the total screening constant of the protons and $\delta B_0$ represents the local deviations from $B_0$ that are not dependent on temperature.

A temperature increase will cause the screening constant of the protons to increase linearly and result in a decrease of the local magnetic field. A GRE sequence can be used to acquire phase distribution images (phase maps) of the phantom. By acquiring another phase map of the same slice after heating and subtracting the images, the change in phase caused by heating can be acquired. The Larmor equation shows that the phase (radians) measured within a voxel at a temperature $T$ is given by (Eq. 9)

$$\varphi(T) = \gamma TE[(1 - (T))B_0 + \delta B_o] \quad (9)$$

where $\gamma$ is the gyromagnetic ratio of hydrogen (42.577 x 10[6] MHz) and $TE$ is the echo time (ms). The change in phase caused by an increase of temperature to $T'$ is determined by (Eq. 10)

$$\Delta\varphi = \varphi(T') - \varphi(T) = \gamma TE[\sigma_{total}(T) - \sigma_{total}(T')]B_0 = -\gamma TE\alpha\Delta TB_0 \quad (10)$$

where $\alpha$ is the temperature-dependent chemical shift coefficient (ppm/°C), a constant in the linear temperature dependence of $\sigma_{total}$ [23,24]. Calibration experiments of the temperature dependence of the water PRF shift in tissues suggest values of 0.9 to 1.1 x 10[-8]/°C [21], where the average temperature-dependent chemical shift coefficient for pure water is approximately 1 x $10^{-8}$/°C from -15°C to 100°C [25].

Once reordered (Eq. 11) it can be used to determine temperature maps of the phantom,

$$\Delta T = \frac{\Delta \varphi}{\alpha \gamma B_0 TE} \quad (11)$$

# 3. MATERIALS AND METHODS

## 3.1 Preparation of phantom

A $T_1$ doped agar-gel phantom was created by dissolving agar (60 g/L), NaCl (10 g/L) and copper sulphate $CuSO_4$ (1 g/L) in distilled hot water. Copper sulphate was used as a preservative and a dopant to increase the signal-to-noise (SNR) ratio by decreasing the $T_1$ relaxation time. By reducing the $T_1$ relaxation time, a larger fraction of the spin population will be allowed to return to its original spin state in the same amount of time [26]. To ensure the solution was homogeneous, it was stirred for 15 minutes with a magnetic stirrer at room temperature. The solution was then cured for 15 minutes at 121 °C in a pressure-controlled chamber under 150 kPa (1.5 bar), to minimise residual bubbles. Finally, the phantom was convectively cooled down to room temperature.

To minimise thermal effects, a 3 L round bottomed, borosilicate glass (with a specific heat capacity of 80 J/kg.K [27]) flask was used as a phantom container. The phantom's specific heat capacity of 4292 J/kg.K was experimentally determined with a NETZSCH DSC 214 differential scanning calorimeter.
The $T_1$ properties of the phantom were determined at room temperature using a STIR sequence over 50-20000 ms.

## 3.2 PRF thermometry method validation

The PRF method for estimating SAR was first verified by comparing SAR values from two independent measures:

- whole body calorimetry: Thermal imaging,
- site specific: Fibre optics.

The phantom temperature was measured using external no-contact calorimetric infrared thermometry. This method was used to give an estimate of the location of hot spots during heating. A FLIR ONE Pro thermal camera with 70 mK thermal sensitivity was pointed down the bore of the scanner (Siemens Symphony 1.5 T). Heating along the edges of the phantom was observed, suggesting the phantom was being heated by SAR.

A two-channel OTG-M170 fiber optic temperature sensor connected to a TempSens multi-channel signal conditioner (Opsens Solutions Ltd) with ± 0.3 K total system accuracy was incorporated into the phantom. The two-fibre optics were embedded, one at the centre of the phantom and the second at the periphery, 10 cm from the surface and perpendicular to centre of the phantom. This allowed for discrete points of measurement in the phantom. The recorded temperature shift values of 0.5 K and 2.3 K at the centre and periphery (see Figure 1), respectively were used to validate the PRF method at discrete points.

After the results of the PRF method (see Table 2) were shown to agree with the independent measures and scanner readout, only the PRF method was used in further measurements.

## 3.3 Independent SAR assessment using PRF thermometry

A temperature map was obtained by phase difference mapping. The phantom was inserted into the head coil and placed inside the bore overnight to achieve thermal equilibrium before starting the experiment. A baseline image was acquired using a 2D fast gradient echo sequence read right to left and then left to right (parameters described in Table 1). The difference of these images was calculated to create a phase map. This method allows for the rapid creation of phase maps of the phantom before and after heating; the short acquisition time means there is minimal heating/cooling of the phantom to increase accuracy.

GRE parameters 2 NEX; 256 × 256 matrix; 16 × 16 cm FOV; 2 mm slice thickness; 30° flip angle; 7.0 ms echo time (TE**);** and 46 ms repetition time (TR). Duration 11 s.

A high-power 2D fluid-attenuated inversion recovery (FLAIR) sequence was used to exaggerate the effects of RF induced heating. This sequence was chosen as it is a commonly used sequence and deposits a large amount of RF energy in a short period of time. The heating time was varied for different scanners to give a range of SAR values. The heating parameters used for each scanner are summarized in Table 1.

| FLAIR sequence | Siemens 1.5 T Symphony | Philips 3 T Achieva | GE 1.5 T Signa Explorer | GE 1.5 T Signa Explorer | Siemens 1.5 T Magnetom |
|---|---|---|---|---|---|
| Acquisition type | 2D | 2D | 2D | 2D | 2D |
| Phase-encoding direction | AP | AP | AP | AP | AP |
| Slice thickness (mm) | 0.9x0.9x5 | 0.9x0.9x5 | 2x2x6 | 1.8x1.8x6 | 1x1x5 |
| Repetition time (ms) | 8760 | 12000 | 9000 | 9000 | 8760 |
| Echo time (ms) | 104 | 118 | 87 | 87 | 104 |
| Matrix size | 256x256 | 256x256 | 256x256 | 256x256 | 256x256 |
| Inversion time (ms) | 2500 | 2850 | 2500 | 2500 | 2500 |
| Spacing between slices (mm) | 0 | 3.6 | 5 | 5 | 0 |
| Echo train per slice | 18 | 23 | 6 | 7 | 15 |

| Scan time (min:sec) | 15:32 | 3:02 | 2:26 | 3:12 | 12:13 |
|---|---|---|---|---|---|
| SAR reported (W/kg) | 1.88 | 1.52 | 0.42 | 0.55 | 1.50 |

Table 1 Heating parameters

The phase mapping method was repeated after heating with identical parameters to the baseline GRE acquisition. By subtracting the two images, a field map of the temperature induced phase shift was generated. A temperature map was subsequently created by applying equation 11 to the image. By averaging the temperature changes over the whole phantom an estimate of the global SAR was made using equation 2. The calculated SAR was then compared to the scanner SAR readout.

# 4. RESULTS

## 4.1 PRF thermometry method validation

The results of the SAR values measured by a Siemens 1.5 T Symphony scanner, PRF thermometry and fibre optic probes are summarized in Table 2. The SAR readout of the system was in good agreement with the PRF method. Probe values of 0.5 K and 2.3 K temperature shift at the centre and the periphery of the phantom respectively, validated the PRF heatmap.

| | SAR [W/kg] | Difference |
|---|---|---|
| Siemens 1.5 T Symphony | 1.88 | N/A |
| Proton resonance shift (PRF) | 1.90 | 0.2 % |

Table 2 PRF thermometry method validation - Initial results.

Images of a sample GRE image, phase map and heat map are shown in Fig. 1 marked A, B and C respectively. Even though great care was taken to create a homogeneous phantom, imperfections can be seen in the GRE image. Little information can be gathered from the initial phase map, but after the application of the PRF formula, a heat map can be generated. The heat map shows the magnitude of the phase shift at each pixel, red being the largest phase shift and indigo being the smallest. Heating was highest in the periphery and least in the centre of the phantom, which was expected.

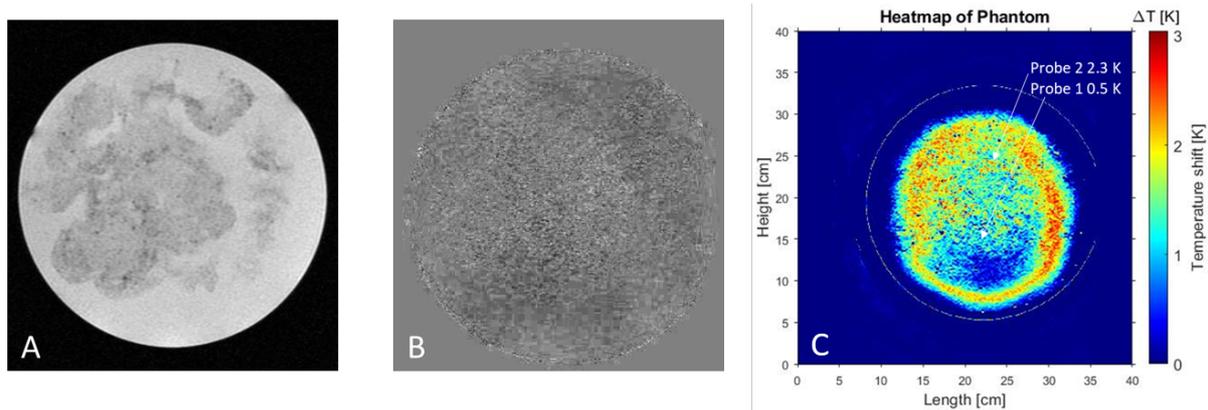

*Figure 1 A. GRE image of phantom B. Phase map of phantom C. Heat map of phantom, note heating around periphery of phantom*

## 4.2 SAR assessment using PRF thermometry

After validating the PRF method, four more scanners were tested with the results summarized in Table 3.

| MRI Model | Scanner readout [W/kg] | Calculation [W/kg] | Difference |
|---|---|---|---|
| Siemens 1.5 T Symphony | 1.88 | 1.90 | 1.0 % |
| Philips 3 T Achieva | 1.52 | 1.52 | 0 % |
| GE 1.5 T Signa Explorer | 0.42 | 0.41 | -2.3 % |
| GE 1.5 T Signa Explorer | 0.55 | 0.56 | 1.7 % |
| Siemens 1.5 T Magnetom Sola | 1.50 | 1.49 | -0.7 % |

Table 3 SAR measurement results using PRF method

The experimentally calculated SAR values were in good agreement with the manufacturer's values. For all scanners the percentage difference ranged from 0 to 2.3%. For the Siemens 1.5 T Symphony MRI system, the MRI scanner-reported SAR value was 1.88 W/kg and the calculated SAR value was 1.90 W/kg. For the Philips 3 T Achieva MRI system, the MRI scanner-reported SAR value was 1.52 and the calculated SAR value was 1.52 W/kg. For the two same GE 1.5 T Signa Explorer MRI systems, the calculated SAR values were 0.42 and 0.55 W/kg and the calculated SAR values were 0.41 and 0.56 W/kg respectively. For the Siemens 1.5 T Magnetom Sola MRI system, the MRI scanner-reported SAR value was 1.50 W/kg and the calculated SAR value was 1.49 W/kg. The percentage differences between the measured and reported SAR values for the Siemens 1.5 T Symphony, Philips 3 T Achieva, both GE 1.5 T Signa Explorer's and Siemens 1.5 T Magnetom Sola were 1.0%, 0%, -2.3%, 1.7% and -0.7%, respectively.

# 5. Discussion

The goal of this study was to develop a method that could be used in annual medical physics QA's to verify the SAR output of an MRI scanner. While numerical models are faster to implement and don't require direct scanner access, it is important for medical physicists to make independent measurements experimentally. The PRF method showed good agreement with both our validation methods and the scanner readouts, while the initial results seem promising with small errors (maximum 2.3%) compared to the MRI-scanner reported values. We would like to stress that these are not absolute measurements.

This method is quick to compute using our in-house MATLAB software, and has the advantage of not requiring any external equipment such as fibre optics. The method was shown to work for a variety of different scanner models and field strengths of both 1.5 and 3 T. As mentioned previously, SAR has an approximate quadratic dependence on the field strength. As a result, future studies at higher field strengths at 7 T are planned.

While the phantom can be created within 24 hours, degradation of the phantom is still an issue. Dehydration over time will result in less accurate results. We plan to investigate this further and improve our method in future projects.

Seo et al. reported less accurate results using a plastic capsule phantom. They performed thermometry using DTI rather than GRE [9]. We believe the use of a glass rather than plastic container (which has a lower specific heat capacity (80 J/kg.K) compared to 1700-1900 J/kg.K for plastics) gave improved accuracy. The use of GRE imaging rather than DTI reduces the phantom cooling time (GRE 11 s, DTI 4-5 min) between scans, giving more accurate results. However, field inhomogeneity will have a greater adverse effect on GRE acquisitions and must be considered when performing this technique.

Differences in SAR distributions between homogeneous phantoms and heterogeneous human anatomies have been documented [28] and necessitate exploring more detailed head phantoms. Tissue heterogeneity is usually more defined in numerical head models [29-31]. It has been shown that local SAR values are dependent on tissue geometry and heterogeneity, especially at higher field strengths [29]. Hotspots may also occur in close proximity to RF coil capacitors due to the capacitive coupling of the electric field or around the metal components of implants. A more precise method in the future could help to identify these hotspots to further improve medical physics QA's.

This PRF method is highly susceptible to movement. As previously discussed, the temperature maps obtained with this method are created by calculating the phase difference between a baseline image and an image acquired after heating. Any motion between the acquisition of these images will lead to the occurrence of artefacts. As a result, this method is particularly difficult to implement in-vivo. Methods have been created to try and reduce these artefacts. With respect to respiration, respiratory gating has been implemented, but can fail when the respiratory cycle is irregular [32,33]. Another approach is to remove the need for a baseline image; these methods are called *referenceless or self-referenced thermometry* and attempt to generate temperature estimates from each individual image without a preheating reference scan. [34,35]. This method required heating an area that is at least partially surrounded by a region that hasn't been heated with an adequate SNR. This method could be explored further for in-vivo methods.

Phase unwrapping can be an issue when generating phase maps due to magnetic field inhomogeneities [36]. This is due to the phase being calculated by the tangent inverse function that applies modulo $2\pi$ operation to the true phase. This results in the calculated phase being limited, or 'wrapped', to a range of ($-\pi$, $\pi$), leading to discontinuities appearing in the phase function [37]. Phase unwrapping algorithms [38,39] aim to remove these artificial phase jumps, though these can still cause problems when the image contains deformation or severe noise. A water phantom was used before testing to check for magnetic field inhomogeneities. If severe inhomogeneities were detected, the test was not continued as the scanner would fail a standard QA protocol.

# 6. Conclusions

We have successfully created a protocol to independently verify the SAR output of an MRI scanner. Experimental values were in good agreement with the manufacturers' values for the five scanners tested. Currently SAR is not routinely measured in annual medical physics quality assurance checks. As MRI field strengths increase, the need for routine testing and validation of SAR levels is ever greater. The method proposed in this work could be used to provide an independent annual validation of manufacturers SAR values. More work is still required to develop a non-degrading phantom and to verify this method at higher field strengths.

# Acknowledgements

The project was co-financed by the European Regional Development Fund (ERDF) under Ireland's European Structural, Investment Funds Programme 2014-2020 and Enterprise Ireland; Grant agreement: CF-2017-0826-P, Irish Research Council postgraduate scholarship GOIPG/2018/82 and the NUI Galway College of Science.
Thank you to the NUI Galway Microbiology and physics department for their help creating the phantom. Authors would like to thank Maja Drapiewska and David Connolly from the NUI Galway College of Engineering and Informatics and Eileen Smith from Netzsch for help with DSC measurements.

14